\documentclass[pre,aps,twocolumn,superscriptaddress,amsmath,showpacs,floatfix]{revtex4}
\usepackage{graphicx}
\begin{document} 
 \title{Elasticity of a system with non-central potentials}
 \author{Michael Murat}
 \altaffiliation{Permanent address: Soreq Nuclear Research Center, Yavne 81800, Israel}
 \email{michael@soreq.gov.il}
 \affiliation{School of Physics and
 Astronomy, Raymond and Beverly Sackler Faculty of Exact Sciences,
 Tel Aviv University, Tel Aviv 69978, Israel}
 \author{Yacov Kantor}
 \email{kantor@post.tau.ac.il}
 \affiliation{School of Physics and
 Astronomy, Raymond and Beverly Sackler Faculty of Exact Sciences,
 Tel Aviv University, Tel Aviv 69978, Israel}
 
\date{\today}
 
\begin{abstract}
We derive expressions for determination of the stress and the
elastic constants in systems composed of particles interacting via 
non-central two-body potentials as thermal averages of products of
first and second partial derivatives of the interparticle potentials and 
components of the interparticle separation vectors. These results are 
adapted to hard potentials, when the stress and the elastic constants 
are expressed as thermal averages of the components of  normals to 
contact surfaces between the particles and components of vectors 
separating the centers of the particles. The averages require the 
knowledge of simultaneous contact probabilities of two pairs of particles. 
We apply the expressions to particles for which a contact function can 
be defined, and demonstrate the feasibility of the method by computing 
the stress and the elastic constants of a two-dimensional system of 
hard ellipses using Monte Carlo simulations.
\end{abstract}

\maketitle
\section{Introduction}
The mechanical response of materials to deformations is described by
the elasticity theory \cite{lanlif}. The simplest homogeneous (affine) 
deformation of  a continuum can be expressed by a linear dependence 
of the distorted position ${\bf r}$ on the original position of 
that point ${\bf R}$ via relation 
\begin{equation}\label{def:M}
r_i=M_{ij}R_j\ ,
\end{equation}
where $M_{ij}$  is a {\em constant} tensor. We will consider both 
two-dimensional (2D) and three-dimensional (3D) systems, in which
Latin subscripts will indicate Cartesian coordinates. (Summation over
repeated Latin subscripts is implied.) Tensor $M_{ij}$ can be separated 
into an identity tensor $I$ and a non-trivial part
\begin{equation}\label{def:epsilon}
M_{ij}=\delta_{ij}+\epsilon_{ij}\ .
\end{equation}
In general, $\epsilon_{ij}$ can be separated into a {\em symmetric tensor}
representing deformation and an anti-symmetric tensor representing
rotation. We neglect the rotation and assume that  
$\epsilon_{ij}=\epsilon_{ji}$. While the tensor $\epsilon_{ij}$
has a convenient meaning in actual experiments, usually the
elastic deformations are formulated in the terms of the
{\em Lagrangian strain tensor} $\eta_{ij}$, which defines the
change in the distances between the points \cite{etausage}:
\begin{equation}\label{def:eta}
r_ir_i=M_{ik}M_{il}R_kR_l\equiv(\delta_{kl}+2\eta_{kl})R_kR_l\ .
\end{equation}

In the case of affine deformations the definitions in 
Eqs. \ref{def:M}--\ref{def:eta} are valid for arbitrarily values of
$\epsilon_{ij}$ and $\eta_{ij}$. However, we will further assume that the
deformations are small. For a homogeneous continuum, it suffices
to apply the deformation described by Eq.~\ref{def:M} at the
{\em boundaries} of the system to assure that the same equation
describes every internal point, while in the case of inhomogeneous
system, application of such a deformation to the boundaries assures
that the {\em mean} deformation is equal to $\epsilon_{ij}$ \cite{hashin}. 

Elastic properties of a condensed matter system
describe the energetic cost of a deformation. However, real system
consisting of many moving atoms/molecules cannot be simply represented
by a strain tensor assigned to every point in space. Rather, we can
assume that the {\em boundaries} of such system undergo an affine
deformation described by Eq.~\ref{def:M}. In such a case, the mean 
free energy density, $f$, which is the free energy $F$ divided by the
original (unstrained) volume $V_0$ of the system, can be expanded 
in a power series in the strain variables
\begin{equation}\label{def:C}
f(\{\eta\})=f(\{0\})+\sigma_{ij}\eta_{ij} +\frac{1}{2}
C_{ijkl}\eta_{ij} \eta_{kl}+\ldots
\end{equation}
The coefficients in this expansion are the stress tensor
$\sigma_{ij}$ and the tensor of the (second order) elastic 
constants $C_{ijkl}$, characterizing a given material. 
In the case of isotropic pressure $p$, the stress can be
written as $\sigma_{ij}=-p\delta_{ij}$. Elastic
constants may serve as an indicator of instabilities
associated with phase transitions \cite{birch,zhou}. 

Elastic response of a system to a deformation can be determined
without actually distorting the system, since equilibrium
correlation functions contain all the necessary information.
Indeed almost four decades ago Squire, Holt and Hoover (SHH) \cite{shh}
developed a formalism that extended the theory of elasticity of
Born and Huang \cite{born} to a finite temperature situation
and expressed the elastic properties of a system as thermal
averages of various derivatives of interparticle potentials.
In a certain sense the formalism is an extension of the virial
theorem \cite{virtheorem} which relates the thermal averages
of the products of interparticle forces and the interparticle 
separations to the stress tensor. (Similar formalism enables
evaluation of the elastic properties of 2D membranes in 3D space
\cite{farago_pincus}.) The SHH method is very well 
adapted for use in numerical simulations in constant volume (and 
shape) ensembles. Other methods, extracting the elastic properties
from shape and volume changes of systems, have also been developed
and extensively used \cite{other}.

Usually molecules are not spherically symmetric and we may expect
interactions that depend on the orientation of the molecules. The
introduction of rotational degrees of freedom into the theoretical
description of a system has an interesting effect on the stress and 
elastic constants. At very low densities (almost ideal gas) the
rotational degrees of freedom add a large contribution to the
total kinetic energy of a molecule, but do not contribute to the
pressure. In a thermodynamical treatment of stress we are interested
in the translational degrees of freedom. However, for 
non-spherically-symmetric potentials, the particle rotation may 
have a significant indirect contribution even at moderate densities. 
At larger densities phase diagram may be strongly influenced by
those degrees of freedom.
The general approach of SHH \cite{shh} (see also Refs.
\cite{zhou,bavaud}) was applied in a detailed form to the 
case of particles interacting via central two-body forces.
However, the formalism can be easily extended to the systems of
particles interacting via non-central potentials, as will be shown
in Section \ref{sec:softelast}. 

Modelling of various systems frequently involves particles that 
interact  via {\em hard} potentials which are either 0 or $\infty$: 
in fact simulation of 2D hard disk system dates back to the origins 
of the Metropolis Monte Carlo (MC) method \cite{origMetro}. An obvious 
reason for the use of such potentials in simulations is their numerical 
simplicity. However, there are important physical reasons for such 
models: in many situations entropy plays a dominant role 
in physical processes, and the absence of energy scale in hard
potentials ``brings out'' the entropic features of the behavior.
Hard sphere systems have been the subject of an intensive research
for several decades now (see \cite{gast} and references therein).
They serve as the simplest models for real fluids, glasses, and
colloids. The phase diagram of hard spheres is well known. In
3D this system undergoes an entropically driven first-order phase 
transition from liquid to solid phase \cite{hsfreezing}. Elastic 
constants of such solids have been explored in the past 
\cite{numres,runge}. Entropy also plays a crucial role in the 
systems containing long polymers, such as gels and rubbers 
\cite{polentropA,degennes_polymer,polentropB}. 
Not surprisingly, hard sphere potentials have been extensively
used to represent excluded volume interactions between the
monomers (see \cite{baum} and references therein).
Kantor {\it et al.} \cite{kkn} introduced {\it tethering 
potential}, that has no
energy but limits the distance between bonded monomers, to
represent covalent bonds in polymeric systems. Such hard potential
combined with hard sphere repulsion can be used to simulate a
variety of polymeric systems.  Recently, Farago and Kantor 
\cite{fk_formalism} adapted the formalism of SHH \cite{shh} to
hard potentials.  This new formalism enabled a study of a sequence 
of entropy-dominated systems, such as 2D \cite{fk_2D} and 3D 
\cite{fk_3D} gels near sol-gel transition and other systems \cite{fk_net}.

Orientation of non-spherically-symmetric molecules plays a
crucial role in the properties of liquid crystals
\cite{degennes_liquid}. For instance, the nematic phase is 
translationally
disordered but it has orientational order of the molecules. From
the early stages of the liquid crystal research it has been
realized that the {\it entropic} part of the free energy related
to non-spherical shapes of the molecules, by itself, can explain
many of the properties of the systems \cite{ons}.  Not
surprisingly, hard potentials were frequently used to investigate
the properties of liquid crystals. Even such simplifications as
infinitely thin disks \cite{frenkel_thindisk} or infinitely thin
rods \cite{frenkel_thinrod} provide valuable insights into the
problem. A slightly more realistic picture is provided by hard
spheroids \cite{fmm}.  Such simulations were primarily motivated
by the desire to understand the liquid phases. However,
two interesting {\it solid phases} have been detected: 
both phases are translationally ordered, but only one
of them has orientational order of spheroids. The orientational 
order is absent only when the spheroid resembles a sphere. 
Sufficiently oblate or prolate spheroids are orientationally 
ordered in the solid phase. During the last twenty
years hard spheroids have be studied in great detail
\cite{spheroids}. A similar hard potential system that is suitable
for the study of the liquid crystals is a collection of hard
spherocylinders (cylinders capped at their ends by hemispheres).
These molecules have slightly more complex phase diagram (which
includes smectic-A liquid-crystalline phase), and also have been
studied in great detail \cite{spherocyl}. Like spheroids, they
have two solid phases. (Spherocylinders do not have a  shape
resembling oblate spheroid.) Taken together, spheroids and
spherocylinders provide a rather coherent picture of influence of
molecular shape on the phase diagram (see, e.g., \cite{sp}). Hard
potentials also have been used in other ways to represent
non-spherically symmetric molecules by combining several spheres
or disks into more complicated shapes, such as heptagons
\cite{woj}, or long rods \cite{hardlong}. To make the models more
realistic, sometimes attractive interaction has been added to the
usual hard repulsive potential \cite{attr}. 

In Section \ref{sec:softelast} we present the formalism for soft
non-central pair potentials. This formalism cannot be directly
applied to the calculation of elastic constants of hard potential
systems. In Section \ref{sec:hardelast} and Appendix 
\ref{sec:regularization} we show how generalization of the approach 
used for centrally-symmetric hard potentials  \cite{fk_formalism} 
can be used to derive expressions applicable to non-central hard 
potentials. In Section \ref{sec:application} we detail the method
by which the formal results can be applied to hard particles for 
which a ``contact function" can be defined. In particular, we 
describe how our formalism can be used for a 2D system of hard 
ellipses. In Section \ref{sec:results} we demonstrate the 
implementation by calculating stress and elastic constants in 
different phases of a system of hard ellipses.

\section{Elastic properties for soft pair potentials}\label{sec:softelast}

In this section we derive explicit expressions for the stress and elastic
constant tensors following the method of SHH \cite{shh} for a more
general case. We will consider  potential energy $\cal V$ which 
can be expressed as
\begin{equation}\label{def:V}
{\cal V}=\sum_{\langle\alpha\beta\rangle}\Phi({\bf r}^{\alpha\beta},\Omega^\alpha,
\Omega^\beta) ,
\end{equation}
where $\Phi$ is the interaction potential of a pair of particles.
Greek indices $\alpha$ and $\beta$ denote particles (atoms/molecules), 
and $\langle\alpha\beta\rangle$ denotes a pair of particles. The above 
equation contains summation over all possible particle pairs. (In this 
paper we do {\em not} assume summation over repeated Greek indices 
indicating particles.) Here 
${\bf r}^{\alpha\beta}={\bf r}^{\beta}-{\bf r}^{\alpha}$ is the vector
connecting two particles, while $\Omega^\alpha$ is the orientation of particle 
$\alpha$. The two-body potential is not necessarily spherically 
symmetric. In fact, we will apply our results to particles that do not 
posses such a symmetry.
We denote all two-body potentials by the same letter $\Phi$ although
nowhere in this formal derivation it is required that they should be 
identical for different pairs of particles.  (It should be
denoted $\Phi^{\alpha\beta}({\bf r}^{\alpha\beta},\Omega^\alpha,
\Omega^\beta)$; however, we omit the superscript of $\Phi$ for brevity.)
From the physical point of view we expect that the potential should be
rotationally invariant, i.e. when the vector ${\bf r}^{\alpha\beta}$ and the orientations of two molecules (described by $\Omega^\alpha$ and $\Omega^\beta$)
perform a ``rigid body'' rotation, the interaction energy should not change.
In fact the symmetry of the stress tensor assumes the presence of
rotational invariance. However, we do not explicitly use this property
in the derivation of the following expressions.

Unlike the central force case \cite{shh} we will need to use both 
$\eta_{ij}$ and $\epsilon_{ij}$ in the process of derivation. Note that 
in the definition of $\eta_{ij}$ in Eq.~\ref{def:eta} only the symmetric
sum $\eta_{ij}+\eta_{ji}$ appears for $i\ne j$. Therefore, without loss 
of generality it is assumed that the Lagrangian strain is a 
{\em symmetric} tensor ($\eta_{ij}=\eta_{ji}$). From Eqs.~\ref{def:epsilon} and \ref{def:eta} we find that
$\eta_{kl}=\frac{1}{2}(\epsilon_{kl}+\epsilon_{lk}+\epsilon_{ik}\epsilon_{il})$.
For small deformations this relation can be inverted to the second order as
\begin{equation}\label{epsiloneta}
\epsilon_{kl}=\eta_{kl}-\frac{1}{2}\eta_{mk}\eta_{ml}+\dots
\end{equation}

In statistical-mechanical description of a solid in a canonical ensemble
we may ask how the free energy $F$ of the solid changes when the
{\em boundaries} of the solid undergo deformation described by Eq.~\ref{def:M}.
In calculation of such $F(\{\eta\})$ we do not impose any restrictions
on the positions or orientations of the particles except the
change in the boundary conditions. The free energy can be expressed via the
partition function as 
\begin{equation}\label{def:F}
F(\{\eta\})=-kT\ln [Z_0Z(\{\eta\})] ,
\end{equation}
where only the configurational part $Z(\{\eta\})$ of the partition
depends on the deformation, while the remaining (``kinetic'') part
$Z_0$ is independent of deformations.  We note, that in classical
physics the details of the inertia tensors of the molecules can
modify the details of their actual motion, but play no role in the
statistical-mechanical properties of the system. Only the
asphericity of the potential matters. The configurational part
\begin{equation}
Z=
\int\limits_{V(\{\eta\})}\prod_{\lambda=1}^Nd{\bf r}^\lambda
\int\prod_{\lambda=1}^Nd\Omega^\lambda\ 
{\rm e}^{-{\cal V}({\bf r}^1,\Omega^1,\dots,
{\bf r}^N,\Omega^N)/kT},
\end{equation}
where ${\bf r}^\alpha$ and $\Omega^\alpha$ represent the position and 
orientation of particle $\alpha$ and $\cal V$ is the interaction potential, 
depends on the deformation only through the distortion of the integration
volume $V(\{\eta\})$ of the possible positions of each of the particles. The
integration over all possible spatial directions of each particle
remains unchanged. If we formally change the integration variable
${\bf r}^\alpha$ for each particle $\alpha$ to the variable ${\bf R}^\alpha$,
which are related by Eq.~\ref{def:M}, then, the limits of interaction
of the new variables will correspond to the undistorted volume $V(\{0\})\equiv V_0$,
and consequently
\begin{equation}\label{def:Z}
Z=
\int\limits_{V_0}\prod_{\lambda=1}^Nd{\bf r}^\lambda
\int\prod_{\lambda=1}^Nd\Omega^\lambda
J{\rm e}^{
-{\cal V}(M_{ij}R^1_j,\Omega^1,\dots,
M_{ij}R^N_j,\Omega^N)/kT},
\end{equation}
where $J$ is the Jacobian corresponding to the change of coordinates
\begin{equation}
J=|\det(M)|^N=|\det(I+2\eta)|^{N/2}\ .
\end{equation}
The deformation now appears as distortion of the coordinates in the
potential $\cal V$.

Thus, the stress and the elastic constants can be viewed as the first
and the second derivatives of the free energy density with respect
to various $\eta_{ij}$. Since in the expansion in Eq.~\ref{def:C} always
appear pairs of terms such as $C_{1123}\eta_{11}\eta_{23}+
C_{1132}\eta_{11}\eta_{32}$
which contain identical $\eta$ terms, we can choose to define
the tensor in a symmetric form $C_{ijkl}=C_{jikl}=C_{ijlk}$.
(An additional symmetry $C_{ijkl}=C_{klij}$ is also evident from
the definition of the tensor.) Strictly speaking, since 
$\eta_{12}=\eta_{21}$ they should be treated as a single variable 
while taking the derivatives
of the free energy density. However, terms containing those two
variables also appear twice in Eq.~\ref{def:C}. Thus, one can simply
treat $\eta_{12}$ and $\eta_{21}$ as independent variables, and symmetrize
the results with the interchange of indices at the end. Alternatively,
one may view each derivative $\partial/\partial \eta_{12}$
as  $\frac{1}{2}(\partial/\partial \eta_{12}+\partial/\partial \eta_{21})$.
Below we always present fully symmetrized expressions.

From Eqs. \ref{def:C} and \ref{def:F} we can express the stress tensor
\begin{equation}
V_0\sigma_{ij}=\frac{\partial{F}}{\partial{\eta_{ij}}}
\biggm|_{\{\eta\}=\{0\}} 
=-\frac{kT}{Z}\frac{\partial{Z}}{\partial{\eta_{ij}}}\biggm|_{\{\eta\}=\{0\}}
\end{equation}
and the second order elastic constants
\begin{eqnarray}
&&V_0C_{ijmn}=\frac{\partial^2 F}{\partial\eta_{mn}\partial\eta_{ij}}
\biggm|_{\{\eta\}=\{0\}} \nonumber\\
&&=\left[\frac{kT}{Z^2}\frac{\partial{Z}}{\partial{\eta_{mn}}}
\frac{\partial{Z}}{\partial{\eta_{ij}}} -
 \frac{kT}{Z}\frac{\partial^2{Z}}{\partial{\eta_{mn}}\partial{\eta_{ij}}}\right]
\biggm|_{\{\eta\}=\{0\}}     
\end{eqnarray}
in terms of the derivatives of $Z$. As can be seen from Eq.~\ref{def:Z}
the dependence of $Z$ on the deformation is contained in the Jacobian $J$
and in the arguments of the potential $\cal V$. The Jacobian depends directly
on $\eta_{ij}$ and its derivatives can be easily calculated. In particular,
we find (see, e.g., Ref. \cite{fk_formalism}) that 
\begin{equation}
\frac{\partial J}{\partial\eta_{ij}}\biggm|_{\{\eta\}=\{0\}}=N\delta_{ij}
\end{equation}
\begin{equation}
\frac{\partial^2 J}{\partial\eta_{mn}\partial\eta_{ij}}\biggm|_{\{\eta\}=\{0\}}=
N^2\delta_{ij}\delta_{mn}-N\delta_{im}\delta_{jn}-N\delta_{in}\delta_{jm}.
\end{equation}
Taking the derivatives of $\cal V$ involves the differentiation of 
the potential with respect to $M_{ij}$, followed by the differentiation 
of $M_{ij}$ with respect to $\epsilon_{kl}$ using Eq.~\ref{def:epsilon}, 
followed by the differentiation of $\epsilon_{kl}$ with respect to 
$\eta_{mn}$ using Eq.~\ref{epsiloneta}. This leads to the following 
expressions for the stress and  elastic constants:
\begin{widetext}
\begin{equation}\label{softstress}
V_0\sigma_{ij}= -NkT\delta_{ij}+\frac{1}{2}\sum_{\langle\alpha\beta\rangle}
\left\langle \frac{\partial\Phi\ }{\partial R^{\alpha\beta}_i}R^{\alpha\beta}_j+
 \frac{\partial\Phi\ }{\partial R^{\alpha\beta}_j}R^{\alpha\beta}_i\right\rangle\ ,
\end{equation}
\begin{eqnarray}\label{softelast}
&&V_0C_{ijmn}=NkT(\delta_{im}\delta_{jn}+\delta_{in}\delta_{jm})\nonumber\\
&+&\frac{1}{4kT}
\sum_{\langle\alpha\beta\rangle}\left\langle 
\frac{\partial\Phi}{\partial R^{\alpha\beta}_i}R^{\alpha\beta}_j+
\frac{\partial\Phi}{\partial R^{\alpha\beta}_j}R^{\alpha\beta}_i
\right\rangle
\sum_{\langle\gamma\delta\rangle}\left\langle 
\frac{\partial\Phi}{\partial R^{\gamma\delta}_m}R^{\gamma\delta}_n+
\frac{\partial\Phi}{\partial R^{\gamma\delta}_n}R^{\gamma\delta}_m
\right\rangle  \nonumber\\
&-&\frac{1}{4kT}\sum_{\langle\alpha\beta\rangle,\langle\gamma\delta\rangle}
\Bigg\langle \frac{\partial\Phi}{\partial R^{\alpha\beta}_m}
\frac{\partial\Phi}{\partial R^{\gamma\delta}_i} R^{\alpha\beta}_n  R^{\gamma\delta}_j
+\frac{\partial\Phi}{\partial R^{\alpha\beta}_n}
\frac{\partial\Phi}{\partial R^{\gamma\delta}_i} R^{\alpha\beta}_m  R^{\gamma\delta}_j
+\frac{\partial\Phi}{\partial R^{\alpha\beta}_m}
\frac{\partial\Phi}{\partial R^{\gamma\delta}_j} R^{\alpha\beta}_n  R^{\gamma\delta}_i
+\frac{\partial\Phi}{\partial R^{\alpha\beta}_n}
\frac{\partial\Phi}{\partial R^{\gamma\delta}_j} R^{\alpha\beta}_m  R^{\gamma\delta}_i
\Bigg\rangle
 \nonumber\\
&+&\frac{1}{4}\sum_{\langle\alpha\beta\rangle}\Bigg\langle
\frac{\partial^2\Phi}{\partial R^{\alpha\beta}_m\partial R^{\alpha\beta}_i}R^{\alpha\beta}_nR^{\alpha\beta}_j
+\frac{\partial^2\Phi}{\partial R^{\alpha\beta}_n\partial R^{\alpha\beta}_i}R^{\alpha\beta}_mR^{\alpha\beta}_j
+\frac{\partial^2\Phi}{\partial R^{\alpha\beta}_m\partial R^{\alpha\beta}_j}R^{\alpha\beta}_nR^{\alpha\beta}_i
+\frac{\partial^2\Phi}{\partial R^{\alpha\beta}_n\partial R^{\alpha\beta}_j}R^{\alpha\beta}_mR^{\alpha\beta}_i  \Bigg\rangle
 \nonumber\\
&-&\frac{1}{8}\sum_{\langle\alpha\beta\rangle}\Bigg\{
\left\langle \frac{\partial\Phi}{\partial R^{\alpha\beta}_j}R^{\alpha\beta}_n+
\frac{\partial\Phi}{\partial R^{\alpha\beta}_n}R^{\alpha\beta}_j \right\rangle\delta_{im}
+\left\langle \frac{\partial\Phi}{\partial R^{\alpha\beta}_i}R^{\alpha\beta}_m+
\frac{\partial\Phi}{\partial R^{\alpha\beta}_m}R^{\alpha\beta}_i \right\rangle\delta_{jn}
\nonumber\\
&+&\left\langle \frac{\partial\Phi}{\partial R^{\alpha\beta}_i}R^{\alpha\beta}_n+
\frac{\partial\Phi}{\partial R^{\alpha\beta}_n}R^{\alpha\beta}_i \right\rangle\delta_{jm}
+\left\langle \frac{\partial\Phi}{\partial R^{\alpha\beta}_j}R^{\alpha\beta}_m+
\frac{\partial\Phi}{\partial R^{\alpha\beta}_m}R^{\alpha\beta}_j \right\rangle\delta_{in} \Bigg\} \ .
\end{eqnarray}
\end{widetext}
In the above equations we already use the coordinates $\bf R^{\alpha\beta}$
of the undistorted system to emphasize the fact that all the averages are now
calculated in the absence of the deformation.

Since $(\partial\Phi/\partial R^{\alpha\beta}_i)R^{\alpha\beta}_j=
-f^{\alpha\beta}_iR^{\alpha\beta}_j$, where ${\bf f}^{\alpha\beta}$ is
the force between the particles $\alpha$ and $\beta$, we can recognize in 
Eq.~\ref{softstress} the standard virial theorem, although in the textbooks \cite{virtheorem} the derivation is quickly reduced to calculation 
of the (isotropic) pressure $p$.

The accuracy of the expressions \ref{softstress} and \ref{softelast} can 
be verified by reducing the above formulae to the case of isotropic 
central force potential in which case
\begin{equation}
\frac{\partial\Phi}{\partial R_i} = \Phi'\frac{R_i}{R}\ ,
\end{equation}
where prime denotes a derivative of $\Phi$ with respect to 
inter-particle separation $R$. Similarly,
\begin{equation}
\frac{\partial^2\Phi}{\partial R_i\partial R_j}=
\Phi''\frac{R_iR_j}{R^2}
+\Phi'\frac{\delta_{ij}}{R}
-\Phi'\frac{R_iR_j}{R^3}
\end{equation}
After performing these substitutions, we recover the standard expressions 
for central force potentials \cite{shh}.

\section{Hard potentials}\label{sec:hardelast}

\begin{figure}
\includegraphics[height=5cm]{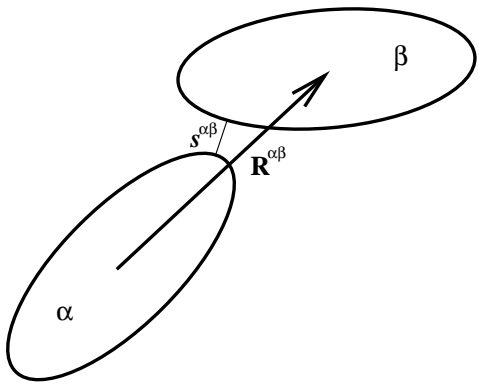}
\caption{\label{fig:eldefs} Two ``hard'' particles at a close approach.
${\bf R}^{\alpha\beta}$ is a vector connecting their centers, and
$s^{\alpha\beta}$ is the minimal separation of the surfaces.
 }
\end{figure}

The expressions for stress and elastic constants that have been obtained 
in the previous section, presumed smooth potentials with well defined
first and second derivatives. There is a certain difficulty in 
translating the expressions obtained for soft potentials to a hard 
potential situation. For instance, the term
$(\partial\Phi/\partial R^{\alpha\beta}_i)R^{\alpha\beta}_j$
in Eq.~\ref{softstress} looks poorly defined for a hard potential that
changes on contact between 0 and $\infty$. However, we note that the derivative
of the potential really originates from the derivative
$\partial[{\rm e}^{-\Phi({\bf R}^{\alpha\beta},\Omega^\alpha,\Omega^\beta)/kT}]/
\partial R^{\alpha\beta}_i$. The latter, is a derivative of a step function
that changes between 0 and 1, when the potential changes between $\infty$
and 0. This observation, has been used in Refs. \cite{runge,barker} to derive
simple expressions for the pressure of hard sphere system. A detailed 
and rigorous description of the various aspects (including calculation of 
pressure) of interaction of hard  convex solids can be found in Ref. \cite{hardpotreview}. A typical interaction between two  hard particles 
(e.g., ellipses in 2D or ellipsoids in 3D) is represented by
Fig. \ref{fig:eldefs} which depicts a close approach of two such particles, 
when ${\bf R}^{\alpha\beta}$ is the vector connecting their centers, 
while $s^{\alpha\beta}$ denotes the minimal distance between them. 
Gradient of  
${\rm e}^{-\Phi({\bf R}^{\alpha\beta},\Omega^\alpha,\Omega^\beta)/kT}$
with respect to ${\bf R}^{\alpha\beta}$, when $\Omega^\alpha$ and 
$\Omega^\beta$ are kept fixed, is simply 
${\bf n}^{\alpha\beta}\delta(s^{\alpha\beta})$
where ${\bf n}^{\alpha\beta}$ is unit vector perpendicular to the 
surfaces at the point of contact, pointing from $\alpha$ to $\beta$.
Thus we can make the substitution
\begin{eqnarray}\label{substitution}
&&\left\langle \frac{\partial\Phi}{\partial R^{\alpha\beta}_i}R^{\alpha\beta}_j\right\rangle
\rightarrow -kT\int d{\bf R}^\alpha d{\bf R}^\beta d\Omega^\alpha d\Omega^\beta\times
\nonumber\\
&&
n^{\alpha\beta}_i\delta(s^{\alpha\beta})
R^{\alpha\beta}_jP({\bf R}^\alpha,\Omega^\alpha,{\bf R}^\beta, \Omega^\beta)
\end{eqnarray}
where
\begin{eqnarray}\label{eq:Pdef}
&&P({\bf R}^\alpha,\Omega^\alpha,{\bf R}^\beta, \Omega^\beta)=
\nonumber\\
&&\frac{1}{Z}\int\prod_{\lambda=1\atop\lambda\ne\alpha,\beta}^N(d{\bf R}^\lambda d\Omega^\lambda){\rm e}^{-\sum_{\langle\mu\nu\rangle\atop\ne\langle\alpha\beta\rangle}\Phi({\bf R}^{\mu\nu},\Omega^\mu,\Omega^\nu)/kT}
\end{eqnarray}
is the probability density of particles $\alpha$ and $\beta$ to be at particular
positions and orientations. The integral in Eq.~\ref{substitution} simply
represents the average of $n^{\alpha\beta}_iR^{\alpha\beta}_j$ over all possible
contacts between two particles weighted with proper probability densities of
those contacts. (Since the probability density changes from a finite value,
when the particles are almost in contact, to zero, when the particles overlap,
we need to consider the case when $s^{\alpha\beta}$ approaches 0 from the positive
side.) Thus, if we denote 
$\Delta_i^{\alpha\beta}=n^{\alpha\beta}_i\delta(s^{\alpha\beta}-0^+)$, we can
find the hard potential limit by replacement
\begin{equation}\label{hardlim}
\frac{\partial\Phi}{\partial R^{\alpha\beta}_i}
\rightarrow -kT\Delta_i^{\alpha\beta}
\end{equation}
By substituting the result into Eq.~\ref{softstress} we obtain the following
expression for the stress in the case of hard potentials:
\begin{equation}\label{eq:hardstress}
\frac{V_0\sigma_{ij}}{kT}= -N\delta_{ij}-\frac{1}{2}\sum_{\langle\alpha\beta\rangle}
\left\langle \Delta_i^{\alpha\beta}R^{\alpha\beta}_j+
 \Delta_j^{\alpha\beta}R^{\alpha\beta}_i\right\rangle .
\end{equation}

The substitution appearing in Eq.~\ref{hardlim} cannot be used to calculate
the elastic constants in Eq.~\ref{softelast}, since the latter expression
contains a double summation (over pairs $\langle\alpha\beta\rangle$ and
$\langle\gamma\delta\rangle$) of the derivatives of the potentials. Such summation
includes a term where $\langle\alpha\beta\rangle=\langle\gamma\delta\rangle$.
A direct substitution of Eq.~\ref{hardlim} would lead to the appearance
of the term $[\delta(s^{\alpha\beta})]^2$ which causes the expression to diverge.
(This is a {\em true divergence}, rather than some mathematical subtlety of
approaching the limit of hard potentials.) However, Eq.~\ref{softelast} also
contains the second derivative of $\Phi$, which becomes poorly defined in the
hard potential limit. We shall show in the Appendix \ref{sec:regularization}
that the {\em sum} of the apparently divergent and poorly defined terms has a well defined hard 
potential limit. In fact this sum can be transformed into an expression 
in Eq.~\ref{eq:Chard} which includes only first derivatives of potentials 
and products of the first derivatives of potentials of {\em different} 
particle pairs. While the resulting expression in Eq.~\ref{eq:Chard} 
looks more complicated, it can be simply transformed  (using 
Eq.~\ref{hardlim}) into an expression for the elastic constants of hard particles:

\begin{widetext}
\begin{eqnarray}\label{eq:hardelast}
\frac{V_0C_{ijmn}}{kT}&=&N(\delta_{im}\delta_{jn}+\delta_{in}\delta_{jm})
\nonumber\\
&+&\frac{1}{4}
\sum_{\langle\alpha\beta\rangle}\left\langle 
\Delta_i^{\alpha\beta}R^{\alpha\beta}_j+
\Delta_j^{\alpha\beta}R^{\alpha\beta}_i
\right\rangle
\sum_{\langle\gamma\delta\rangle}\Big\langle 
\Delta_m^{\gamma\delta}R^{\gamma\delta}_n+
\Delta_n^{\gamma\delta}R^{\gamma\delta}_m
\Big\rangle  \nonumber\\
&-&\frac{1}{4}\sum_{\langle\alpha\beta\rangle}\sum_{\langle\gamma\delta\rangle
\atop \ne \langle\alpha\beta\rangle}
\left\langle \Delta_m^{\alpha\beta}
\Delta_i^{\gamma\delta} R^{\alpha\beta}_n  R^{\gamma\delta}_j
+\Delta_n^{\alpha\beta}
\Delta_i^{\gamma\delta} R^{\alpha\beta}_m  R^{\gamma\delta}_j
+\Delta_m^{\alpha\beta}
\Delta_j^{\gamma\delta} R^{\alpha\beta}_n  R^{\gamma\delta}_i
+\Delta_n^{\alpha\beta}
\Delta_j^{\gamma\delta} R^{\alpha\beta}_m  R^{\gamma\delta}_i
\right\rangle
 \nonumber\\
&+&\frac{1}{8}\sum_{\langle\alpha\beta\rangle}
\sum_{\gamma\atop\ne\alpha,\beta}\Big\langle
\Delta_m^{\alpha\beta}
\left(\Delta_i^{\gamma\beta}+\Delta_i^{\alpha\gamma}\right)
R^{\alpha\beta}_nR^{\alpha\beta}_j
+\Delta_n^{\alpha\beta}
\left(\Delta_i^{\gamma\beta}+\Delta_i^{\alpha\gamma}\right)
R^{\alpha\beta}_mR^{\alpha\beta}_j
\nonumber\\
&+&\Delta_m^{\alpha\beta}
\left(\Delta_j^{\gamma\beta}
+\Delta_j^{\alpha\gamma}\right)
R^{\alpha\beta}_nR^{\alpha\beta}_i
+\Delta_n^{\alpha\beta}
\left(\Delta_j^{\gamma\beta}
+\Delta_j^{\alpha\gamma}\right)
R^{\alpha\beta}_mR^{\alpha\beta}_i\Big\rangle
\nonumber\\
&+&\frac{1}{4}\sum_{\langle\alpha\beta\rangle}\Big\{
\left\langle \Delta_j^{\alpha\beta}R^{\alpha\beta}_n+
\Delta_n^{\alpha\beta}R^{\alpha\beta}_j \right\rangle\delta_{im}
+\left\langle \Delta_i^{\alpha\beta}R^{\alpha\beta}_m+
\Delta_m^{\alpha\beta}R^{\alpha\beta}_i \right\rangle\delta_{jn}
+\left\langle \Delta_i^{\alpha\beta}R^{\alpha\beta}_n+
\Delta_n^{\alpha\beta}R^{\alpha\beta}_i \right\rangle\delta_{jm}
\nonumber\\
&+&\left\langle \Delta_j^{\alpha\beta}R^{\alpha\beta}_m+
\Delta_m^{\alpha\beta}R^{\alpha\beta}_j \right\rangle\delta_{in} 
+\left\langle \Delta_m^{\alpha\beta}R^{\alpha\beta}_n+
\Delta_n^{\alpha\beta}R^{\alpha\beta}_m \right\rangle\delta_{ij}
+\left\langle \Delta_i^{\alpha\beta}R^{\alpha\beta}_j+
\Delta_j^{\alpha\beta}R^{\alpha\beta}_i \right\rangle\delta_{mn}
\Big\} .
\end{eqnarray}

\end{widetext}
Each of the terms in the above equation corresponds to some particular
case of contact between pairs of particles. E.g., term of the type
$\langle \Delta_i^{\alpha\beta}R^{\alpha\beta}_j\rangle$ in the
last sum corresponds to a contact between a pair of
particles $\alpha$ and $\beta$, and consequently the contribution of
that sum is proportional to $N$. The sum with the prefactor $\frac{1}{8}$
contains terms corresponding to three touching particles: E.g., the
term $\langle\Delta_m^{\alpha\beta}\Delta_i^{\gamma\beta}R^{\alpha\beta}_n
R^{\alpha\beta}_j\rangle$
corresponds to the situation when particle $\beta$ touches particles $\alpha$ and
$\gamma$ simultaneously. The number of such contacts at any given moment 
is also
proportional to $N$. The second and the third lines of the equation
have $N^2$ different averages corresponding to the number
of pairs of contacts that might appear in the system. However, we note 
that in both sums taken together we always have differences of the
type 
$\langle\Delta_i^{\alpha\beta}R^{\alpha\beta}_j\rangle \langle\Delta_m^{\gamma\delta}R^{\gamma\delta}_n\rangle-
\langle\Delta_i^{\alpha\beta}R^{\alpha\beta}_j \Delta_m^{\gamma\delta}R^{\gamma\delta}_n\rangle$. This term
vanishes if the contact between the pair $\langle\alpha\beta\rangle$
is {\em uncorrelated} with the contact between the pair
$\langle\gamma\delta\rangle$. This happens when the two
pairs are outside the correlation distance. Consequently, only 
pairs of contacts close to each other will contribute, and therefore
the total contribution of those terms is proportional to $N$.

\section{Application to hard ellipse system}\label{sec:application}

Equations \ref{eq:hardstress} and \ref{eq:hardelast} enable calculation
of the stress and elastic constants of a system consisting of hard particles,
provided we are able to calculate thermal averages of the type
$\langle \Delta_i^{\alpha\beta}R^{\alpha\beta}_j\rangle$. This expression
depends on the probability density of particles being in contact.
For the calculation of stress we need only the probability {\em density}
of a single contact between a pair of particles, while the calculation
of elastic constants involves the probability density of two such contacts
happening simultaneously. It is natural to use MC method
\cite{hoover_bk,frenkel_bk,binder_bk} to evaluate averages of this type.
MC simulation of hard potentials is particularly simple since no energy
scale is present, and every elementary move of a particle is either accepted
without a need to calculate the Boltzmann factor, or results in a forbidden
configuration, and is, consequently, rejected. Application, of these
procedures requires a method to identify intersection of two hard particles.
Such  methods have been found both for 2D ellipses \cite{vieillard}
and for 3D ellipsoids \cite{hard_spheroids_funct,perram}.
In Appendix \ref{sec:overlap} we explain in detail the case of ellipses which is
used in the current work. Here, we will consider a slightly more general case
when a simple function can be defined that identifies the contact between
two particles.

\begin{figure}
\includegraphics[height=10cm]{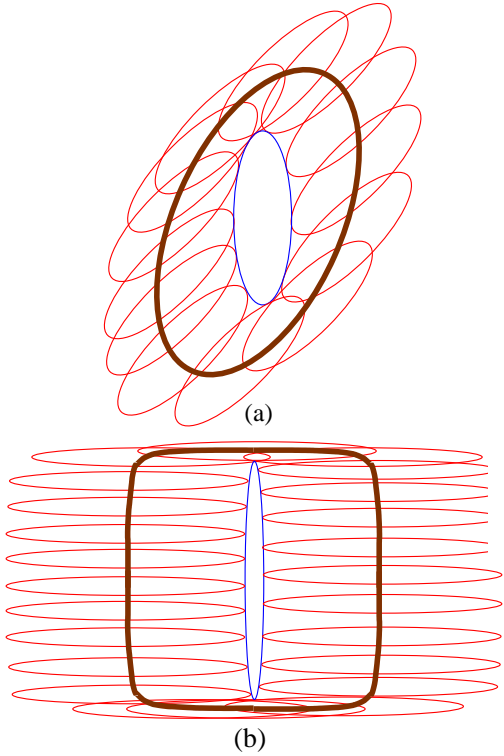}
\vskip 1cm
\caption{\label{fig:elcontact} (a) Center of slightly elongated
ellipse ($a/b=3$) tilted by 45$^o$ with respect to an identical (vertical)
ellipse circumscribes an ``oval'' trajectory, depicted by the thick line,
when the contact point moves along the ellipse.
(b) Center of strongly elongated ellipse ($a/b=14$)
rotated by 90$^o$ with respect to an identical (vertical) ellipse
circumscribes  a ``rounded square'' trajectory, depicted by the thick line,
when the contact point moves along the ellipse.
 }
\end{figure}

Consider a function $\Psi$ that depends on the positions and orientations
of two particles, that vanishes when the particles touch each other and 
is positive when the particles are separated. (The 
formalism can be trivially generalized to the case when the the function
has some other non-vanishing constant value at the contact.) Consider a case
when the orientations of both particles are fixed, and we explore positions
where the function vanishes. Figs.~\ref{fig:elcontact}a and 
\ref{fig:elcontact}b
depict 2D cases of one fixed ellipse, while another (identical) ellipse 
is rotated by a fixed angle and is shown in a variety  positions where 
they touch. The ratio between the major and minor semi-axes of 
the ellipses, $a$ and $b$, respectively, are different in both pictures.
The thick line traces the possible positions of the center of the 
moving ellipse. Note that the shape of such a contact line depends on
the degree of elongation of the ellipses and on their relative orientation.
Those are the positions that are relevant for the calculation of the average 
$\langle \Delta_i^{\alpha\beta}R^{\alpha\beta}_j\rangle$. In 2D
this is a line, while in 3D this is a surface.

\begin{figure}
\includegraphics[height=5cm]{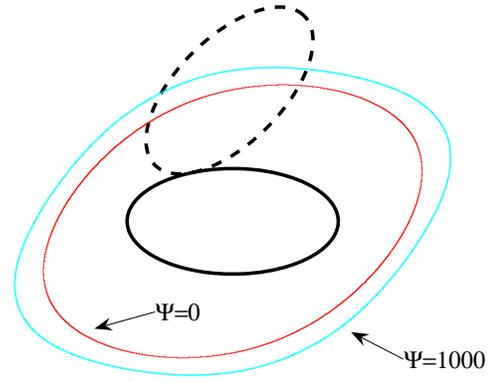}
\caption{\label{fig:psi0} Line of $\Psi=0$ and $\Psi=1000$ for a pair
of ellipses with aspect ratio $E=a/b=2$ with their axes rotated
by 45$^o$ for a function defined in Appendix \ref{sec:overlap}.
Note that the thickness of the area between two lines
of fixed $\Psi$ varies slightly.
 }
\end{figure}

The direction of the force, normal to the contact plane, is also normal 
to this surface. Thus, assuming that $\Psi$ is a sufficiently smooth 
function, we can calculate $n^{\alpha\beta}_i=
(\partial\Psi^{\alpha\beta}/\partial R^{\alpha\beta}_i)/
|\nabla\Psi^{\alpha\beta}|$, where the gradient (and the partial derivative)
with respect to ${\bf R}^{\alpha\beta}$ is taken when the orientations of
the particles are fixed. In thermal average we need to calculate
\begin{eqnarray}\label{eq:example_av}
&&\langle \Delta_i^{\alpha\beta}R^{\alpha\beta}_j\rangle
=\int d\Omega^{\alpha}d\Omega^{\beta}\times
\nonumber\\
&&
\int dS^{\alpha\beta} n^{\alpha\beta}_iR^{\alpha\beta}_j
P({\bf R}^{\alpha},\Omega^{\alpha},{\bf R}^{\beta},\Omega^{\beta}),
\end{eqnarray}
which involves the integration along the contact surface (or line) 
$S^{\alpha\beta}$ of the probability density $P({\bf R}^{\alpha},
\Omega^{\alpha},{\bf R}^{\beta},\Omega^{\beta})$, defined in 
Eq.~\ref{eq:Pdef}, of two particles being in those positions and 
orientations. During a MC simulation such an event strictly never 
occurs. We can replace the integration along the surface, by
an integration inside a thin shell of thickness $t$ along the
contact surface. In such a case $dS^{\alpha\beta}P\approx(dV/t)P=dp/t$,
where $dV$ is the volume element and $dp$ is the {\em probability} 
of the center of a particle
being within a shell, at some particular area. Note, that the thickness
of the shell does not have to be constant, but can vary from place to place
on the contact surface. In fact, we can define the shell as corresponding
to all positions for which $0\le\Psi^{\alpha\beta}<\Psi_0$, where
$\Psi_0$ is some fixed number. Fig. \ref{fig:psi0} depicts such a shell
corresponding to two values of $\Psi^{\alpha\beta}$, for
a function defined in Appendix \ref{sec:overlap}. If $\Psi_0$ is small enough,
we can determine the local thickness of the shell from the relation
$\Psi_0\approx|\nabla\Psi^{\alpha\beta}|t$. Substituting the values
of $t$ and $n^{\alpha\beta}_i$ expressed via function $\Psi^{\alpha\beta}$
into Eq.~\ref{eq:example_av}, and noting that approximate expressions
mentioned in this paragraph become exact for vanishing $\Psi_0$, we
arrive at the expression
\begin{equation}
\langle \Delta_i^{\alpha\beta}R^{\alpha\beta}_j\rangle
=\lim_{\Psi_0\to 0}\frac{1}{\Psi_0}\int d\Omega^{\alpha}d\Omega^{\beta}
\int\limits_{S(\Psi_0)} \frac{\partial\Psi^{\alpha\beta}}{\partial R^{\alpha\beta}_i}R^{\alpha\beta}_j dp
\end{equation}
\begin{equation}\label{eq:lim}
\langle \Delta_i^{\alpha\beta}R^{\alpha\beta}_j\rangle
=\lim_{\Psi_0\to 0} \frac{1}{\Psi_0}\left\langle\frac{\partial\Psi^{\alpha\beta}}{\partial R^{\alpha\beta}_i}R^{\alpha\beta}_j\right\rangle_{S(\Psi_0)}
\end{equation}
In these equation $S(\Psi_0)$ in the integral and in the thermal 
average denote a shell defined by the limit $\Psi_0$ on the function $\Psi^{\alpha\beta}$.

In the numerical calculation of the stress, the limit in 
Eq.~\ref{eq:lim} is not easy to implement: Using a large $\Psi_0$
leads to an inaccurate answer, while using a small $\Psi_0$ leads
to a small number of ``almost contact" events, and, consequently, to
large statistical errors. One may try considering a numerical
extrapolation to $\Psi_0=0$ by measuring the stress for a 
sequence of decreasing $\Psi_0$s. However, the events for
smaller values of $\Psi_0$ are also contained in set of the
events for larger $\Psi_0$s. It is difficult to extrapolate
such correlated sets of data points. The independence of the
data points can be achieved by calculating a sequence of values
of the stress for ``contact shells'' defined by $\Psi_{\alpha\beta}$
located in a sequence of segments $[0,\Psi_0),[\Psi_0,2\Psi_0),\dots
[K\Psi_0,(K+1)\Psi_0),\dots$ ($K$ is integer). Values of $\sigma_{ij}$ 
now can be conveniently extrapolated to their ``real" values. 
A similar method has been used by Farago and Kantor \cite{fk_formalism} 
to calculate the stress and elastic constants of hard sphere solids.

The terms in the expressions for elastic constants including two
pairs of particles can be similarly handled. One simply uses $\Psi_{01}$
and $\Psi_{02}$ to define {\em two} shells, $S(\Psi_{01})$ and 
$S(\Psi_{02})$, respectively, each corresponding to a  particular contact, 
and considers the events which occur when both pairs of particles are 
within their respective shells simultaneously. E.g.,
\begin{eqnarray}
&&\langle \Delta_m^{\alpha\beta}R^{\alpha\beta}_n 
\Delta_i^{\gamma\delta}  R^{\gamma\delta}_j\rangle=
\nonumber\\
\lim_{{\Psi_{01}\to 0\atop\Psi_{02}\to 0}} && \frac{1}{\Psi_{01}\Psi_{02}}\left\langle\frac{\partial\Psi^{\alpha\beta}}{\partial R^{\alpha\beta}_m}R^{\alpha\beta}_n
\frac{\partial\Psi^{\gamma\delta}}{\partial R^{\gamma\delta}_i}R^{\gamma\delta}_j
\right\rangle_{S(\Psi_{01}),S(\Psi_{02})} .
\end{eqnarray}
Compared with the case of the stress, the numerical evaluation of 
the limit where the thickness of the shells vanishes presents an
even bigger numerical problem, since
the probability of two contact events is very small. Nevertheless, this can
be handled in a similar way, by considering a 2D array of segments
of the type  $\{[K\Psi_0,(K+1)\Psi_0),[L\Psi_0,(L+1)\Psi_0)\}$  
($K$ and $L$ integers)
and obtaining the values of the various parts of the elastic constants by
extrapolating the 2D surface to its ``real" value of vanishing contact layer
thickness.

\section{Results of simulations}\label{sec:results}

We used the method developed in this work to calculate the elastic 
properties of 2D hard ellipse system as a part of a study
of its phase diagram \cite{mk}. Here, we briefly demonstrate 
the usefulness of the method. As in any hard particle system,
temperature plays no role, since the interactions have no 
``energy scale." The temperature appears only as a multiplicative
prefactor in the free energy $F$ and in Eq.~\ref{eq:hardstress}
for the stress and Eq.~\ref{eq:hardelast} for the elastic constants. The
results depend on the density of the particles and their size and shape:
we characterized the system by the number of particles per unit
area $\rho$ and by the sizes of the major and minor semi-axes, $a$ 
and $b$, of the ellipses. Frequently, reduced density 
$\rho^*\equiv 4\rho a b$ is used. The maximal possible (close packed) 
$\rho^*$ is  independent of the aspect ratio $E=a/b$ of the ellipses and is
equal to $2/\sqrt{3}\approx 1.155$. It should be noted \cite{vieillard}
that for every fixed $a$ and $b$ there is an infinity of possible 
(equally dense) close 
packed states which are obtained by orienting all ellipses in the same 
direction and packing them into a (distorted triangular) periodic 
structure.

The system of hard disks ($E=1$) has been extensively studied. For
$\rho^*\agt 0.91$ it forms a periodic 2D solid --- a triangular lattice.
The correlation function of atom positions of such a solid decays 
to zero as a power law of the separation between the atoms \cite{mermin}.
Such behavior is usually denoted as quasi-long-range order. At the
same time the orientations of the ``bonds" (imaginary lines connecting 
neighboring atoms) have a long range correlation \cite{mermin_bond}.
The system is liquid for $\rho^*\alt 0.89$, i.e it has no long range 
order of any kind. At the intermediate
densities the system is probably hexatic \cite{hny} --- a phase
with algebraically decaying bond-orientational order, but without
positional order. (However, even very large scale simulations \cite{jm}
have difficulties in distinguishing the hexatic phase from coexisting
solid and liquid phases.) From the point of view of elasticity 
theory, all three phases are isotropic, i.e. their second order elastic
constants are determined by two independent constants. Aspect ratio 
$E\ne1$ of the ellipses adds an additional order parameter ---
their orientation. E.g., for $E=4$ a system of ellipses forms 
isotropic liquid for densities $\rho^*\alt 0.8$. For larger densities 
the ellipses in the liquid become oriented (``nematic phase''). Finally, at
$\rho^*\approx 1.0$ the system becomes a solid of orientationally
ordered ellipses \cite{cuesta}. For weakly elongated ellipses, we
expect the particles to remain orientationally disordered in the entire
liquid phase, and with increasing density to go (possibly via hexatic
phase) to a crystalline state in which the ellipses remain disordered.
The 3D analog of such a state is called {\em plastic
crystal} \cite{plastic}. (Presence of such a phase in almost circular 
ellipses ($E=1.01$) was observed by Vieillard-Baron \cite{vieillard}.)
With increasing density an additional phase transition will bring the 
ellipses into an orientationally ordered state. Phase diagram which 
includes such a transition between two solid phases for 3D hard 
ellipsoids has been studied by Frenkel {\em et al.} \cite{fmm}.

As a test of our formalism we studied a case of moderately elongated
hard ellipses with $E=1.5$. We considered system consisting of 
$N \approx 1000$ ellipses contained in a 2D rectangular box whose 
dimensions were chosen as close as possible to a square. Periodic 
boundary conditions were used.  The ellipses were initially placed 
on a distorted triangular lattice, commensurate with the dimensions
of a closed packed configuration corresponding to this particular
aspect  ratio $E$ and the particular orientation of the ellipses. 
In this Section we describe only the cases when the initial orientation
of the major axes of the ellipses was taken to be perpendicular to 
one of the axes of 
symmetry of the ordered crystal drawn through neighboring particles.
We first performed an equilibration run at constant pressure, in 
order to allow the system to reach an equilibrated state with 
respect to the orientational, as well as the translational ordering. 
The orientational order parameter, the box dimensions and the density 
were monitored during this equilibration run. A MC time unit in the 
equilibration run consisted of $N+1$ elementary moves, one of which, 
on the average, was a volume change attempt, and the rest were particle 
move attempts. A particle move attempt involved choosing a particle 
randomly, and attempting to displace and rotate it simultaneously 
by an amount chosen from a uniform distribution. The move was 
accepted if the displaced particle did not overlap in
its new position and orientation with other particles. The volume 
change attempt was identical to that described in \cite{frenkel_bk}. 
The width of the distributions corresponding to the particle moves 
and the parameter of the volume change were chosen so that the average 
success rates of both types of MC moves were about 50\%.

The use of the constant pressure simulations at the equilibration stage 
enabled us to determine the equilibrium box shape for isotropic 
stress tensor (pressure) conditions. At most densities the final
configuration had orientationally disordered ellipses, and we could
easily verify that the final configurations were independent of the 
specific choice of the starting configuration.
However, at extremely high densities, approaching the close packed
density, even after long equilibration the state resembled the starting
state of orientationally ordered ellipses. Typically, several millions 
of MC time units were required to reach  equilibrium for a given 
pressure. Upon completion of equilibration, we switched to constant 
volume simulations, during which the stresses and the elastic 
constants were evaluated.  Very long simulation times (about 
$10^7$ MC time units) were required for accurate determination of 
these constants, because their calculation depends on extremely 
rare events of two pairs of particles simultaneously touching each 
other. The range of ``contact 
shells'' was chosen in such a way that even in the most remote 
shell the separation between the particles was significantly smaller 
than their mean separation. The statistical accuracy of the elastic 
constants was evaluated by comparing the results of independent runs. 

\begin{figure}
\includegraphics[height=6.5cm]{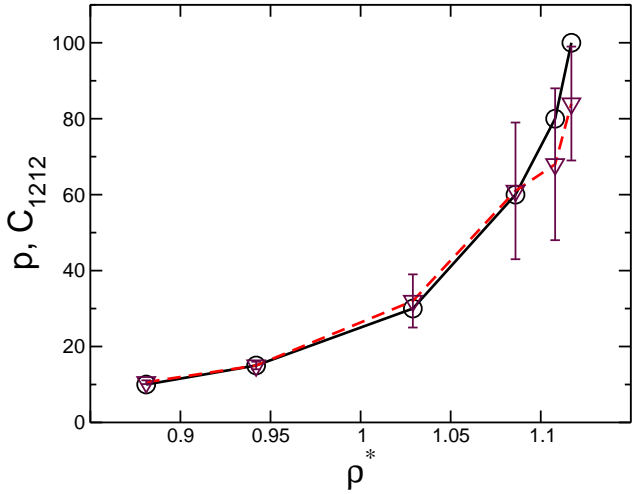}
\caption{\label{fig:elast1} 
Pressure $p$ (open circles connected by solid line) and the elastic constant
$C_{1212}$ (inverted open triangles connected by dashed line) of hard 
ellipse system with aspect ratio $E=1.5$
in the units of $kT/4ab$  as functions of the reduced density $\rho^*$.
The error bars of the pressure are significantly smaller than the
symbols denoting the data points.
 }
\end{figure}
\begin{figure}
\includegraphics[height=6.5cm]{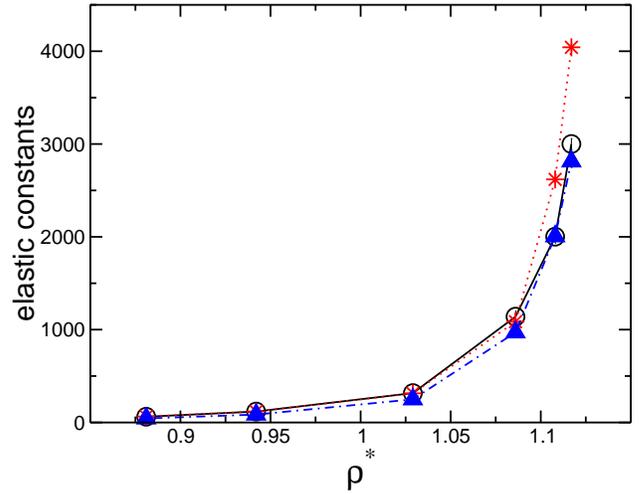}
\caption{\label{fig:elast2} 
Elastic constants $C_{1111}$ (open circles connected by solid line),
$C_{2222}$ (asterisks connected by dotted line) and $C_{1122}$ (full 
triangles connected by dot-dashed line) of hard ellipse system with 
aspect ratio $E=1.5$
in the units of $kT/4ab$  as functions of the reduced density $\rho^*$.
The error bars of all the data points are slightly smaller than the
symbols denoting them.
 }
\end{figure}

\begin{figure*}
\includegraphics[height=6cm]{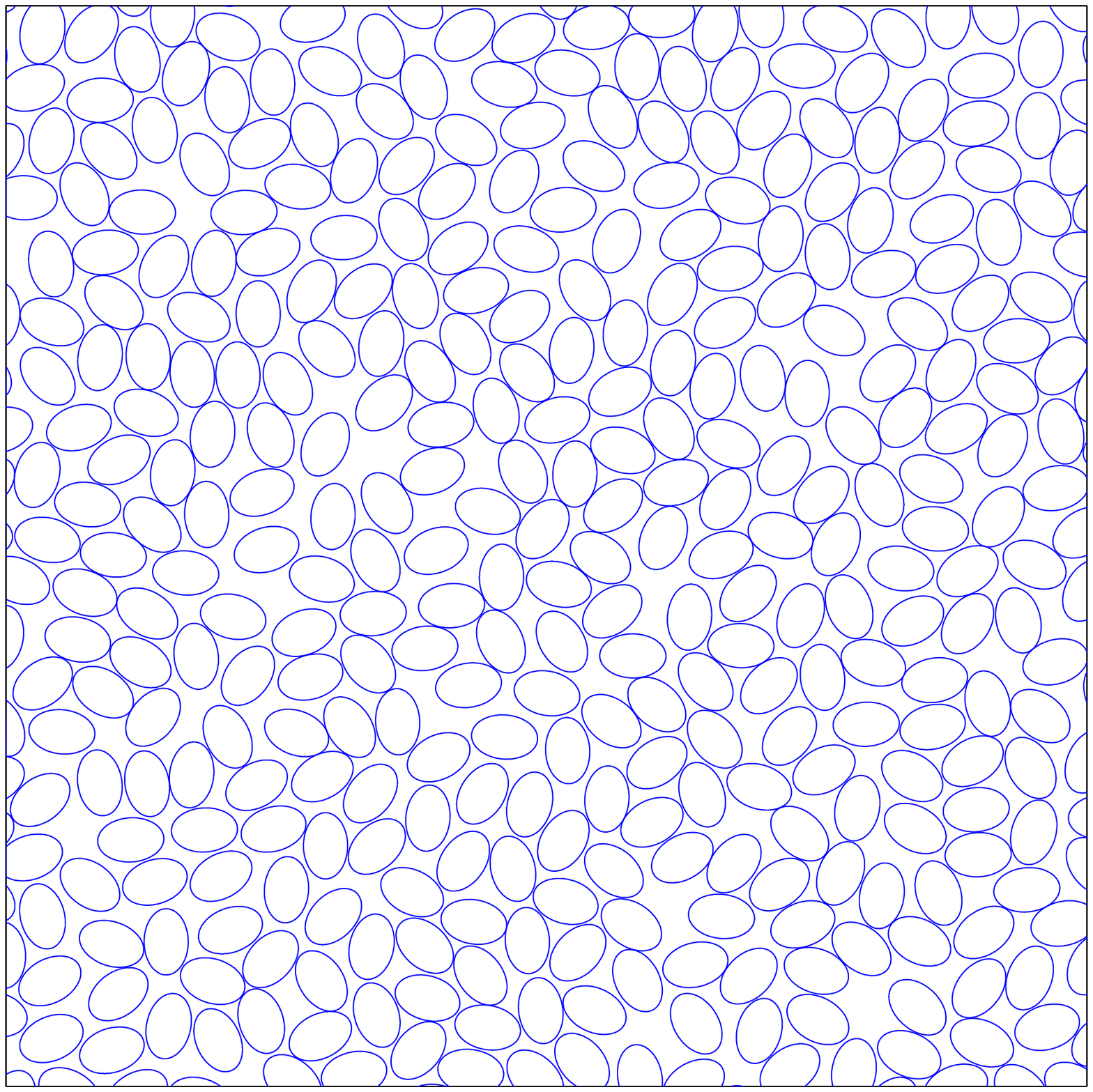}\hskip 3cm
\includegraphics[height=6cm]{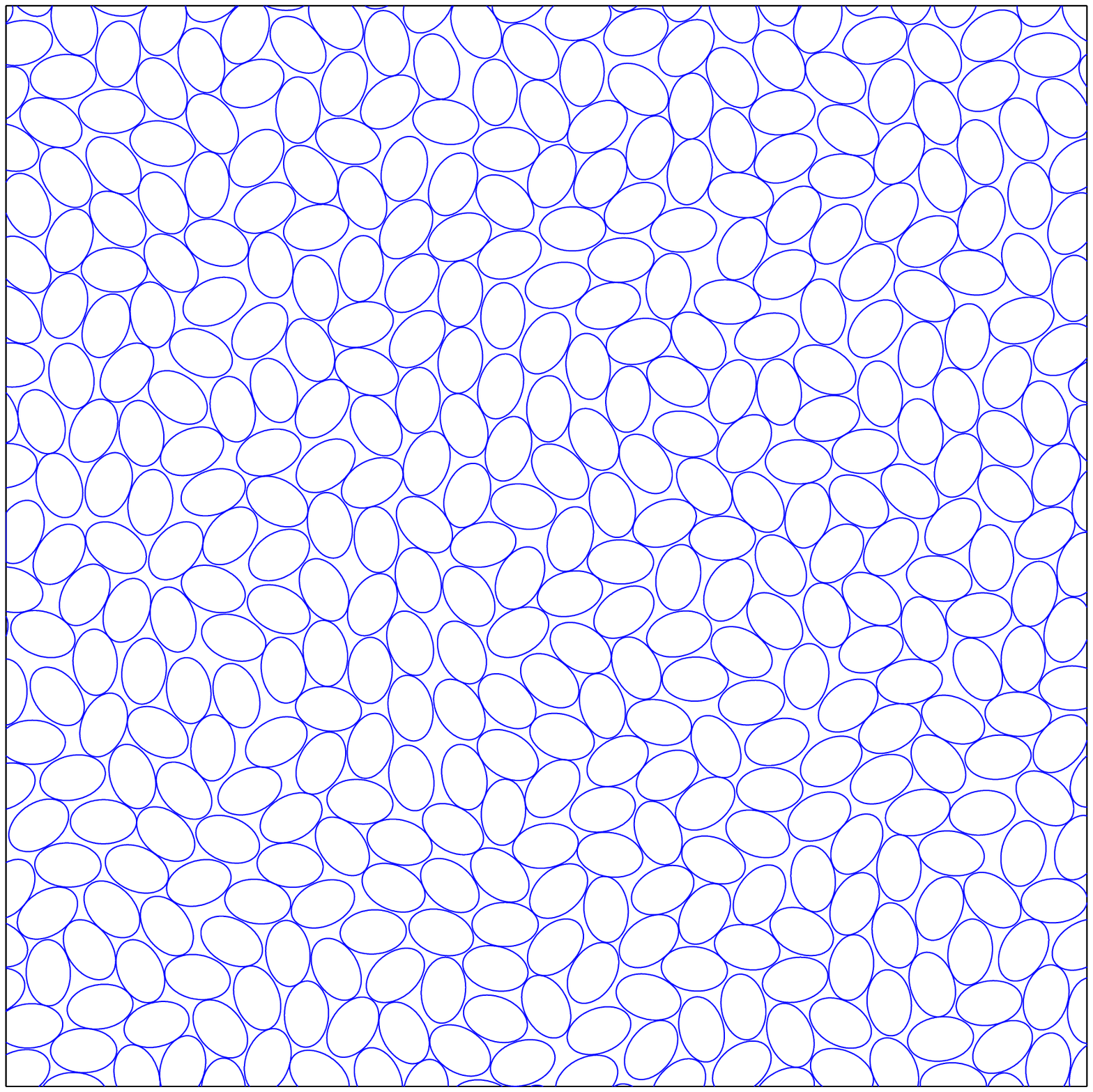}
\par (a){\hskip 8cm}(b)
\par\vskip 1cm
\includegraphics[height=6cm]{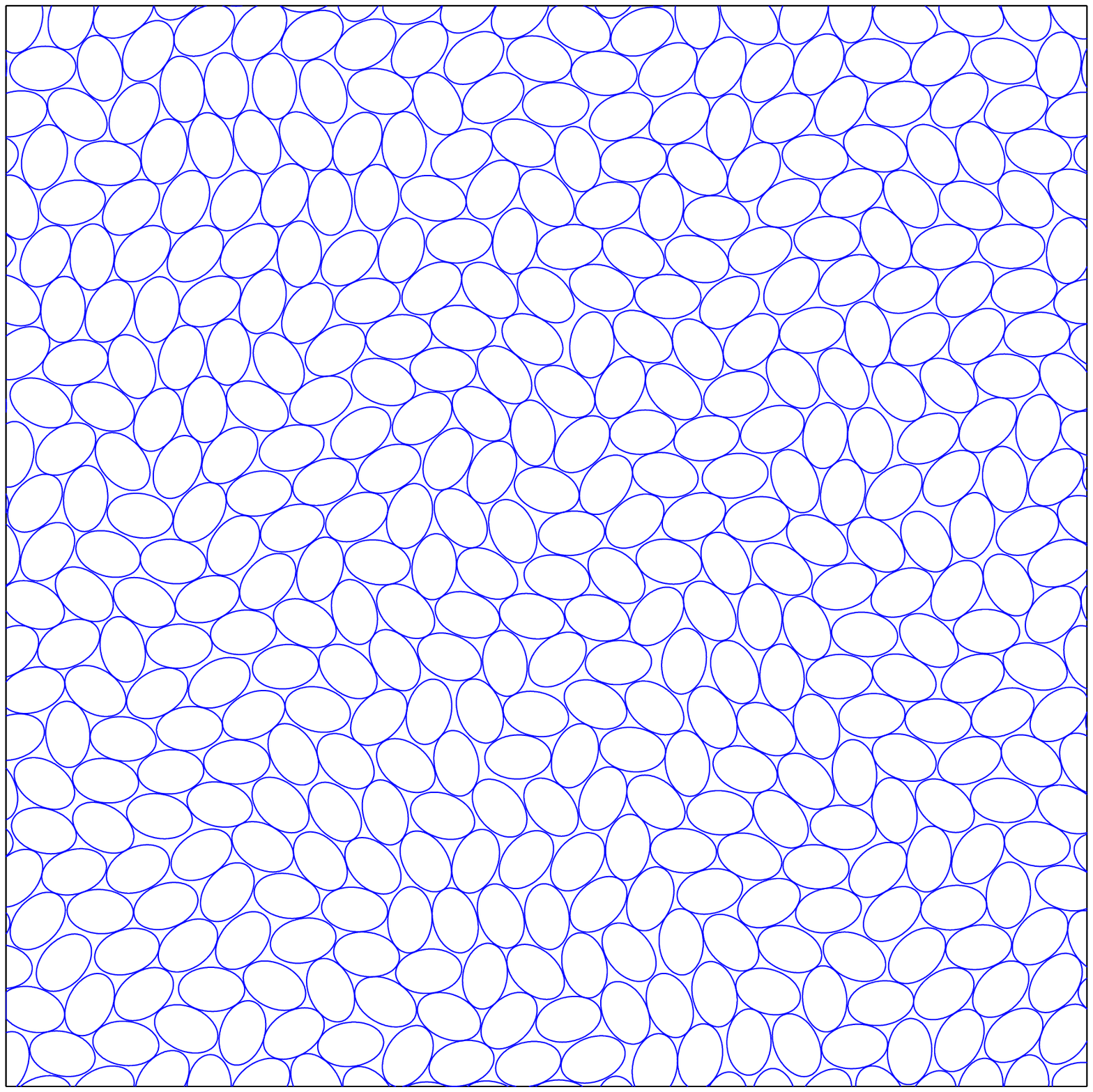}\hskip 3cm
\includegraphics[height=6cm]{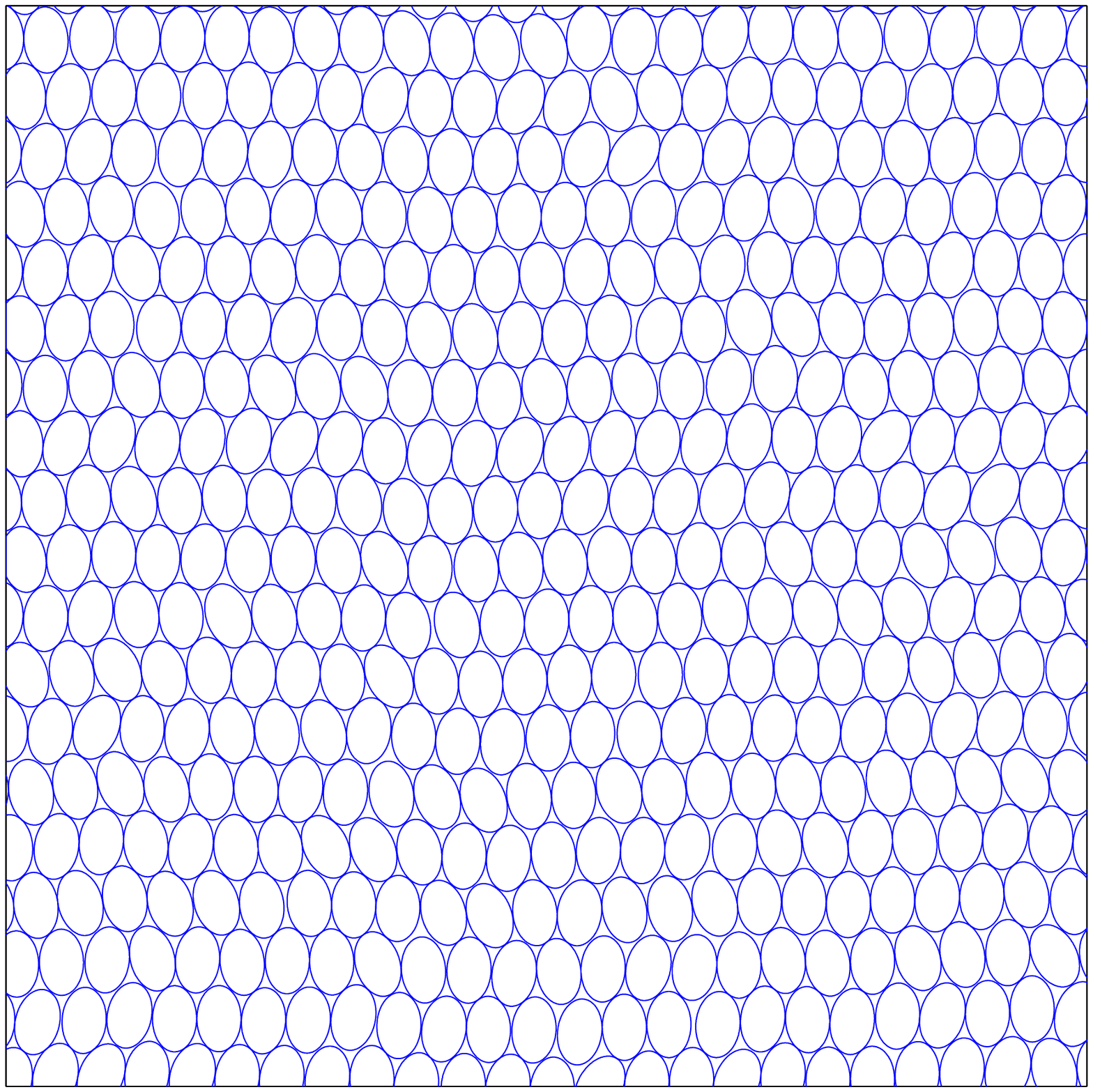}
\par (c){\hskip 8cm}(d)
\caption{\label{fig:configs}
Typical equilibrium configurations of slightly eccentric ($E=1.5$) 
hard ellipse  system  at several densities. Only part of the 
system is shown. All the pictures show the same (partial) volume of
the system; they differ only in the reduced density $\rho^*$:
(a) orientationally and translationally disordered liquid at 
$\rho^*=0.881$;
(b) liquid with a high degree of bond-orientational order at 
$\rho^*=1.029$;
(c) plastic solid with long-range bond-orientational order, 
and quasi-long-range translational order consisting of rotationally 
disordered ellipses at $\rho^*=1.086$;
(d) solid of orientationally ordered ellipses at $\rho^*=1.117$.
 }
\end{figure*}

In a 2D system which has a reflection symmetry with respect to
either $x$ or $y$ axis, the elastic constants with an index 
appearing an odd number 
of times (such as $C_{1112}$) must vanish. Indeed, our simulations
showed that these quantities vanish within the error bars of the
measurement. The system still may have four unrelated elastic constants 
$C_{1111}$, $C_{2222}$, $C_{1122}$ and $C_{1212}$. For a system with
quadratic symmetry the number of independent elastic constants reduces
to three. Such systems are frequently characterized by their bulk
modulus and two shear moduli, $\mu_1=C_{1212}-p$ and 
$\mu_2=\frac{1}{2}(C_{1111}-C_{1122})-p$. For an isotropic system 
$\mu_1=\mu_2$ and, therefore, there are only two independent
constants. (A system with six-fold symmetry is isotropic as far as
the elastic constants are concerned.)
Figs. \ref{fig:elast1} and \ref{fig:elast2} depict the pressure and 
four elastic constants for several values of the reduced density 
$\rho^*$.  One can see that for densities $\rho^* \le 1.086$, 
$C_{1111}$ and $C_{2222}$ practically coincide, indicating that these 
systems are  at least quadratic. Furthermore, the identity 
$C_{1111}-C_{1122}=2C_{1212}$, i.e. $\mu_1=\mu_2$, is found 
to hold within a few percent for these values of $\rho^*$. Consequently,
for these densities the system is isotropic from the point of view of 
the elastic properties. Figs.~\ref{fig:configs}a, b and c depict the
system in that range of densities: Figs.~\ref{fig:configs}a and b 
represent states with vanishing shear moduli, and neither of them
exhibits translational order of the ellipse centers. However, while
Fig.~\ref{fig:configs}a represents a state with all the
characteristics of a liquid, the state in Fig.~\ref{fig:configs}b is 
characterized by slowly decaying bond orientational order, possibly 
indicating hexatic phase. At a slightly higher density, 
Fig.~\ref{fig:configs}c represents a plastic solid: while the ellipses
are randomly oriented, the system exhibits long range bond orientational
order, and algebraically decaying positional order of the particles.
The particles occupy, on the average, positions of an undistorted 
triangular lattice although some undulations are apparent. This is 
characterized by two coinciding positive shear moduli. 

When we approach within few percent the close packed configuration,
corresponding to the two largest densities in Figs.~\ref{fig:elast1} and
\ref{fig:elast2} the prolonged relaxation process does not change the
preferred orientation of the ellipses and the distortion of the
lattice. It would be reasonable to conclude, that at such high
density we finally arrived at the orientationally ordered elastic
solid. The isotropic elastic symmetry no longer holds. We checked
and found that almost all stability criteria \cite{birch,zhou}, 
indicating the sign of the energy change upon small deformation,
are positive. However, one of the shear moduli, namely $\mu_1$,
is slightly negative, although within one standard statistical
deviation from zero. This, may either indicate that we are in
an unstable state, or that there is a continuum of equilibrium states
with different mean orientations of ellipses and, correspondingly,
different dimensions of elementary cell.

\section{Discussion}\label{sec:discussion}

We extended the formalism of SHH \cite{shh} to the case of systems 
interacting
via non-central two-particle potentials. In its form represented by 
Eqs.~\ref{softstress} and \ref{softelast}, the formalism can be used
to study properties of molecular systems. This is particularly
true for highly non-spherical organic molecules, and various soft
condensed matter systems.  The adaptation of the expressions
to hard potentials (Eqs.~\ref{eq:hardstress} and \ref{eq:hardelast})
produced slightly more complicated expressions. However, the 
simplicity of hard potential systems provides excellent insights 
into the entropy-dominated systems. We demonstrated the usefulness 
of the formalism by presenting some results of our study of the hard ellipse
system \cite{mk}. Measurement of several order parameters, and 
the correlation functions is not always sufficient to determine the
nature of phases. For systems of moderate size, it maybe even difficult
to distinguish liquid from a solid. Measurement of elastic constants
provides an additional, very important tool for assessing the nature
of the state of the system. In particular, the elastic constants may
indicate the presence of instability, even when prolonged 
equilibration does not change an existing state. Following the 
indications of instability at high densities, we are currently
performing extensive study of equilibrium states at these densities.

While we worked on a 2D example, the method can be equally well applied 
in 3D for such systems as hard ellipsoids or spherocylinders. 

\acknowledgements 
We would like to thank O. Farago and M. Kardar for useful discussions. 
This research was supported by Israel Science Foundation Grant
No. 193/05.
\bigskip

\appendix

\section{Regularization of diverging and poorly defined terms}\label{sec:regularization}

Section \ref{sec:hardelast} outlines the procedure for transition from
expressions for soft potentials to expressions for hard potentials.
The procedure relies on the fact, that despite the change in the potential
between 0 and $\infty$, the expression for stress and some terms in the
expression for for elastic constants contain a derivative of Boltzmann factor.
Since the latter changes between 1 and 0, its derivatives involve 
$\delta$-function of the distance between the particles, leading to a finite
thermal average.
Even the terms containing contacts between two distinct pairs of particles,
and, consequently, two $\delta$-functions of different variables, produce
a finite results.  However, the third line in Eq.~\ref{softelast}
includes a term containing a product of two derivatives of the same pair
of particles. A substitution of Eq.~\ref{substitution} into such an expression
would lead to appearance of squared $\delta$-function and divergence of the
entire term.   Eq.~\ref{softelast} also
contains second derivative of $\Phi$, which becomes poorly defined in the
hard potential limit. Here we show that {\em sum} the two ``problematic''
terms described above has a well defined hard potential limit.
Let us consider one particle pair $\langle\alpha\beta\rangle$:

\begin{widetext}

\begin{eqnarray}\label{eq:speccase}
&&\Bigg\langle\left(-\frac{1}{kT}\frac{\partial\Phi}{\partial R^{\alpha\beta}_m}
\frac{\partial\Phi}{\partial R^{\alpha\beta}_i}
+\frac{\partial^2\Phi}{\partial R^{\alpha\beta}_m\partial R^{\alpha\beta}_i}\right)R^{\alpha\beta}_nR^{\alpha\beta}_j\Bigg\rangle
\nonumber\\
&&=-kT\int d{\bf R}^\alpha d\Omega^\alpha\int d{\bf R}^\beta d\Omega^\beta
\frac{\partial^2 \ {\rm e}^{-\Phi({\bf R}^{\alpha\beta},\Omega^\alpha,
\Omega^\beta)/kT}}{\partial R^{\alpha\beta}_m\partial R^{\alpha\beta}_i}
R^{\alpha\beta}_nR^{\alpha\beta}_j P({\bf R}^\alpha,\Omega^\alpha,{\bf R}^\beta, \Omega^\beta)
\end{eqnarray}

In the internal integral in Eq.~\ref{eq:speccase} we can replace 
$\partial/\partial R_i^{\alpha\beta}$ by $\partial/\partial R_i^{\beta}$,
since the exponent only depends on ${\bf R}^\beta-{\bf R}^\alpha$. Following
that, we perform integration by parts in which boundary term vanishes and arrive
at
\begin{eqnarray}
&&kT\int d{\bf R}^\alpha d\Omega^\alpha\int d{\bf R}^\beta d\Omega^\beta
\frac{\partial\ {\rm e}^{-\Phi({\bf R}^{\alpha\beta},\Omega^\alpha,
\Omega^\beta)/kT}}{\partial R^{\alpha\beta}_m}
\frac{\partial}{\partial R^{\beta}_i}[
R^{\alpha\beta}_nR^{\alpha\beta}_j P({\bf R}^\alpha,\Omega^\alpha,{\bf R}^\beta, \Omega^\beta)]
\nonumber\\
&&=-\int d{\bf R}^\alpha d\Omega^\alpha\int d{\bf R}^\beta d\Omega^\beta
\frac{\partial \Phi}{\partial R^{\alpha\beta}_m}\ \ {\rm e}^{-\Phi({\bf R}^{\alpha\beta},\Omega^\alpha,
\Omega^\beta)/kT}
 \left(\delta_{in}R^{\alpha\beta}_jP +\delta_{ij}R^{\alpha\beta}_nP
+R^{\alpha\beta}_nR^{\alpha\beta}_j
\frac{\partial P}{\partial R^{\beta}_i}\right).\nonumber\\
\end{eqnarray}
The probability density $P$ in the above expressions was defined in
Eq.~\ref{eq:Pdef}. Variable ${\bf R}^\beta$ appears in every potential
that depends on some ${\bf R}^{\gamma\beta}$, and therefore, the
derivative of $P$ with respect to $R^{\beta}_i$ in the
last term of the above expression can be expressed in the following form:
\begin{eqnarray}
&&\frac{\partial P}{\partial R^{\beta}_i}=
\frac{1}{Z}\int\prod_{\lambda=1\atop\lambda\ne\alpha,\beta}^N(d{\bf R}^\lambda d\Omega^\lambda)\frac{\partial }{\partial R^{\beta}_i}\ \ {\rm e}^{
-\sum_{\langle\mu\nu\rangle\atop\ne\langle\alpha\beta\rangle}\Phi({\bf R}^{\mu\nu},\Omega^\mu,\Omega^\nu)/kT}
\nonumber\\
&=&\frac{1}{Z}\int\prod_{\lambda=1\atop\lambda\ne\alpha,\beta}^N(d{\bf R}^\lambda d\Omega^\lambda)\sum_{\gamma\atop\ne\alpha,\beta}
\frac{\partial }{\partial R^{\gamma\beta}_i}
\ \ {\rm e}^{
-\sum_{\langle\mu\nu\rangle\atop\ne\langle\alpha\beta\rangle}\Phi({\bf R}^{\mu\nu},\Omega^\mu,\Omega^\nu)/kT}
\nonumber\\
&=&-\frac{1}{kTZ}\int\prod_{\lambda=1\atop\lambda\ne\alpha,\beta}^N(d{\bf R}^\lambda d\Omega^\lambda)\sum_{\gamma\atop\ne\alpha,\beta}
\left[\frac{\partial \Phi}{\partial R^{\gamma\beta}_i}\right]
\ \ {\rm e}^{-\sum_{\langle\mu\nu\rangle\atop\ne\langle\alpha\beta\rangle}\Phi({\bf R}^{\mu\nu},\Omega^\mu,\Omega^\nu)/kT}
\end{eqnarray}
Consequently, the term in Eq.~\ref{eq:speccase} becomes:
\begin{equation}
-\left\langle \frac{\partial \Phi}{\partial R^{\alpha\beta}_m}R^{\alpha\beta}_j\delta_{in}
+\frac{\partial \Phi}{\partial R^{\alpha\beta}_m}R^{\alpha\beta}_n\delta_{ij}
\right\rangle
+\frac{1}{kT}\sum_{\gamma\atop\ne\alpha,\beta}\left\langle
\frac{\partial \Phi}{\partial R^{\alpha\beta}_m}\frac{\partial \Phi}{\partial R^{\gamma\beta}_i} R^{\alpha\beta}_n R^{\alpha\beta}_j
\right\rangle
\end{equation}
(This answer is slightly non-symmetric --- this reflects the fact that we
manipulated the integral over the variable ${\bf R}^\beta$, rather that
${\bf R}^\alpha$. In the latter case instead of partial
derivative with respect to $R^{\gamma\beta}_i$ we would have obtained a partial derivative with respect to $R^{\alpha\gamma}_m$. )
When this result is substituted in the expression for the elastic constants in 
Eq.~\ref{softelast} we get
\begin{eqnarray}\label{eq:Chard}
V_0C_{ijmn}&=&NkT(\delta_{im}\delta_{jn}+\delta_{in}\delta_{jm})\nonumber\\
&+&\frac{1}{4kT}
\sum_{\langle\alpha\beta\rangle}\left\langle 
\frac{\partial\Phi}{\partial R^{\alpha\beta}_i}R^{\alpha\beta}_j+
\frac{\partial\Phi}{\partial R^{\alpha\beta}_j}R^{\alpha\beta}_i
\right\rangle
\sum_{\langle\gamma\delta\rangle}\left\langle 
\frac{\partial\Phi}{\partial R^{\gamma\delta}_m}R^{\gamma\delta}_n+
\frac{\partial\Phi}{\partial R^{\gamma\delta}_n}R^{\gamma\delta}_m
\right\rangle  \nonumber\\
&-&\frac{1}{4kT}\sum_{\langle\alpha\beta\rangle}\sum_{\langle\gamma\delta\rangle
\atop \ne \langle\alpha\beta\rangle}
\Bigg\langle \frac{\partial\Phi}{\partial R^{\alpha\beta}_m}
\frac{\partial\Phi}{\partial R^{\gamma\delta}_i} R^{\alpha\beta}_n  R^{\gamma\delta}_j
+\frac{\partial\Phi}{\partial R^{\alpha\beta}_n}
\frac{\partial\Phi}{\partial R^{\gamma\delta}_i} R^{\alpha\beta}_m  R^{\gamma\delta}_j
 \nonumber\\
&+&\frac{\partial\Phi}{\partial R^{\alpha\beta}_m}
\frac{\partial\Phi}{\partial R^{\gamma\delta}_j} R^{\alpha\beta}_n  R^{\gamma\delta}_i
+\frac{\partial\Phi}{\partial R^{\alpha\beta}_n}
\frac{\partial\Phi}{\partial R^{\gamma\delta}_j} R^{\alpha\beta}_m  R^{\gamma\delta}_i
\Bigg\rangle
 \nonumber\\
&+&\frac{1}{8kT}\sum_{\langle\alpha\beta\rangle}
\sum_{\gamma\atop\ne\alpha,\beta}\Bigg\langle
\frac{\partial\Phi}{\partial R^{\alpha\beta}_m}
\left(\frac{\partial\Phi}{\partial R^{\gamma\beta}_i}
+\frac{\partial\Phi}{\partial R^{\alpha\gamma}_i}\right)
R^{\alpha\beta}_nR^{\alpha\beta}_j
+\frac{\partial\Phi}{\partial R^{\alpha\beta}_n}
\left(\frac{\partial\Phi}{\partial R^{\gamma\beta}_i}
+\frac{\partial\Phi}{\partial R^{\alpha\gamma}_i}\right)
R^{\alpha\beta}_mR^{\alpha\beta}_j
\nonumber\\
&+&\frac{\partial\Phi}{\partial R^{\alpha\beta}_m}
\left(\frac{\partial\Phi}{\partial R^{\gamma\beta}_j}
+\frac{\partial\Phi}{\partial R^{\alpha\gamma}_j}\right)
R^{\alpha\beta}_nR^{\alpha\beta}_i
+\frac{\partial\Phi}{\partial R^{\alpha\beta}_n}
\left(\frac{\partial\Phi}{\partial R^{\gamma\beta}_j}
+\frac{\partial\Phi}{\partial R^{\alpha\gamma}_j}\right)
R^{\alpha\beta}_mR^{\alpha\beta}_i\Bigg\rangle
\nonumber\\
&-&\frac{1}{4}\sum_{\langle\alpha\beta\rangle}\Bigg\{
\left\langle \frac{\partial\Phi}{\partial R^{\alpha\beta}_j}R^{\alpha\beta}_n+
\frac{\partial\Phi}{\partial R^{\alpha\beta}_n}R^{\alpha\beta}_j \right\rangle\delta_{im}
+\left\langle \frac{\partial\Phi}{\partial R^{\alpha\beta}_i}R^{\alpha\beta}_m+
\frac{\partial\Phi}{\partial R^{\alpha\beta}_m}R^{\alpha\beta}_i \right\rangle\delta_{jn}
\nonumber\\
&+&\left\langle \frac{\partial\Phi}{\partial R^{\alpha\beta}_i}R^{\alpha\beta}_n+
\frac{\partial\Phi}{\partial R^{\alpha\beta}_n}R^{\alpha\beta}_i \right\rangle\delta_{jm}
+\left\langle \frac{\partial\Phi}{\partial R^{\alpha\beta}_j}R^{\alpha\beta}_m+
\frac{\partial\Phi}{\partial R^{\alpha\beta}_m}R^{\alpha\beta}_j \right\rangle\delta_{in} 
\nonumber\\
&+&\left\langle \frac{\partial\Phi}{\partial R^{\alpha\beta}_m}R^{\alpha\beta}_n+
\frac{\partial\Phi}{\partial R^{\alpha\beta}_n}R^{\alpha\beta}_m \right\rangle\delta_{ij}
+\left\langle \frac{\partial\Phi}{\partial R^{\alpha\beta}_i}R^{\alpha\beta}_j+
\frac{\partial\Phi}{\partial R^{\alpha\beta}_j}R^{\alpha\beta}_i \right\rangle\delta_{mn}
\Bigg\} .
\end{eqnarray}
Since the above expressions contain only the first derivatives of the potentials (or products of such terms for non-identical pairs of particles)
we can use Eqs.~\ref{substitution} and \ref{hardlim} to calculate the elastic 
constants for hard potentials.

\end{widetext}

\section{Overlap function of two ellipses}\label{sec:overlap}

A crucial difficulty in the simulations of hard ellipses in 2D
and hard spheroids in 3D is the determination whether two such 
objects overlap. Vieillard-Baron derived a contact  function for
the 2D case of ellipses \cite{vieillard}. His function
enables a reasonably fast determination of the presence of an overlap.
It can also be used to determine the direction of the inter-ellipse 
force and, consequently, can be used in determination of stresses and
elastic constants. In this Appendix we describe this function and
its properties. 

Consider two identical ellipses, $\alpha$ and $\beta$, whose major 
and minor semi-axes are $a$ and $b$, respectively. Let their centers be separated 
by vector  ${\bf R}^{\alpha\beta}$, and one of them be rotated 
compared to the other one by the angle $\theta$. Projections of 
${\bf R}^{\alpha\beta}$ on the directions of major and minor
semi-axes of ellipse $\alpha$ will be denoted $R^{\alpha\beta}_{\alpha 1}$
and $R^{\alpha\beta}_{\alpha 2}$, respectively. Then we can define a
{\em contact function}
\begin{equation}
\Psi^{\alpha\beta}=4(h_\alpha^2-3h_\beta)(h_\beta^2-3h_\alpha)-(9-h_\alpha h_\beta)^2,
\end{equation}
where for each ellipse we define
\begin{equation}
h_\alpha=1+G-[(R^{\alpha\beta}_{\alpha 1}/a)^2+(R^{\alpha\beta}_{\alpha 2}/b)^2],
\end{equation}
and
\begin{equation}
G=2+\left(\frac{a}{b}-\frac{b}{a}\right)^2\sin^2\theta .
\end{equation}
Function $\Psi^{\alpha\beta}$ depends on the distance between the ellipses and
on their orientation. The number of overlap points of two ellipses can vary 
between 0 and 4. The necessary and sufficient condition for ellipses
to have no intersection points is $\Psi^{\alpha\beta}>0$ and at least
one of the two functions $h_\alpha$ and $h_\beta$ be negative. The proof
of this statement can be found in Ref. \cite{vieillard}. At the contact 
$\Psi^{\alpha\beta}=0$. In the region where the ellipses do not intersect the 
function $\Psi^{\alpha\beta}$ has no extremum points; thus if $\Psi^{\alpha\beta}$
is sufficiently small, i.e.  $0\le\Psi^{\alpha\beta}<\Psi_0$ we can be assured 
that two ellipses are close to each other. 

When the ellipses degenerate into a circles, i.e. $a=b$, the contact function becomes
only a function of the distance between the centers of the circles $R^{\alpha\beta}$
\begin{equation}
\Psi^{\alpha\beta}=3\left(\frac{R^{\alpha\beta}}{a}\right)^6
\left[\left(\frac{R^{\alpha\beta}}{a}\right)^2-4\right] .
\end{equation}
Note, that the gradient of this function at the distance of contact between
the circles has a rather large value $768/a$, so even if we choose $\Psi^{\alpha\beta}=8$ the circles will be about 1\% of the radius away 
from each other.

The shell between the lines $\Psi^{\alpha\beta}=0$ and  $\Psi^{\alpha\beta}=\Psi_0$
has a uniform thickness for circles, but its thickness varies with position in
ellipses as can be seen in the Fig. \ref{fig:psi0}. When the aspect ratio of the 
ellipse becomes much larger than unity the thickness becomes strongly variable 
which will adversely influence the accuracy of Monte Carlo simulations.

\end{document}